\documentclass[sigconf, nonacm]{acmart}

\usepackage{amsmath}
\usepackage{algorithmic}
\usepackage{graphicx}
\usepackage{textcomp}
\usepackage{xcolor}
\usepackage{url}
\usepackage{graphicx}
\usepackage{subcaption}
\usepackage{enumitem}
\usepackage{amsthm}
\usepackage{xcolor}
\usepackage{listings}

\usepackage[a-1b]{pdfx}

\definecolor{bluekeywords}{rgb}{0.13,0.13,1}
\definecolor{greenkeywords}{rgb}{0,0.5,0}
\definecolor{turqusnumbers}{rgb}{0.17,0.57,0.69}
\definecolor{redstrings}{rgb}{0.5,0,0}

\lstdefinelanguage{demoQL}
    {
    keywordstyle=[1]\color{bluekeywords}, 
    keywordstyle= [2]\color{greenkeywords},
    keywordstyle= [3]\bfseries,
    keywords = [1]{QUERY, FROM, AS, TO, LET, BE, IN},
    keywords = [2]{if, then, else, map, any, elem, sum},
    keywords = [3]{cons, nil},
    sensitive=false,
    morestring=[b]",
    stringstyle=\color{redstrings}
    }

\lstnewenvironment{fslisting}
  {
    \lstset{
        language=demoQL,
        basicstyle=\ttfamily,
        breaklines=true,
        columns=fullflexible}
  }
  {
  }

\theoremstyle{definition}
\newtheorem{definition}{Definition}[section]
\newtheorem{example}{Example}[section]

\begin{document}
\title{MultiCategory: Multi-model Query Processing Meets Category Theory and Functional Programming}


\author{Valter Uotila}
\author{Jiaheng Lu}
\affiliation{%
\institution{University of Helsinki}}
\email{first.last@helsinki.fi}

\author{Dieter Gawlick}
\author{Zhen Hua Liu}
\author{Souripriya Das}
\affiliation{%
\institution{Oracle Corporation}}
\email{first.last@oracle.com}

\author{Gregory Pogossiants}
\affiliation{%
\institution{SATS Technologies}}
\email{gregp_21@yahoo.com}

\begin{abstract}
The variety of data is one of the important issues in the era of Big Data. The data are naturally organized in different formats and models, including structured data, semi-structured data, and unstructured data. Prior research has envisioned an approach to abstract multi-model data with a schema category and an instance category by using category theory. In this paper, we demonstrate a system, called \texttt{MultiCategory}, which processes multi-model queries based on category theory and functional programming. This demo is centered around four main scenarios to show a  tangible system. First, we show how to build a schema category and an instance category by loading different models of data, including relational, XML, key-value, and graph data. Second, we show a few examples of query processing by using the functional programming language Haskell. Third, we demo the flexible outputs with different models of data for the same input query.  Fourth, to better understand the category theoretical structure behind the queries, we offer a variety of graphical hooks to explore and visualize queries as graphs with respect to the schema category, as well as the query processing procedure with Haskell.
\end{abstract}

\maketitle



\section{Introduction}
The variety of data is one of the most important issues in modern data management systems to cope with the challenge of Big Data. In many applications,  data sources are naturally organized in different formats and models, including structured data, semi-structured data, and unstructured data. To address the challenge of variety, multi-model databases have begun to emerge with a single database platform to manage multi-model data together, with a fully integrated backend to handle the demands for performance and scalability \cite{journals/csur/LuH19}. 

\begin{figure}[t]
    \centering
    \includegraphics[scale = 0.60]{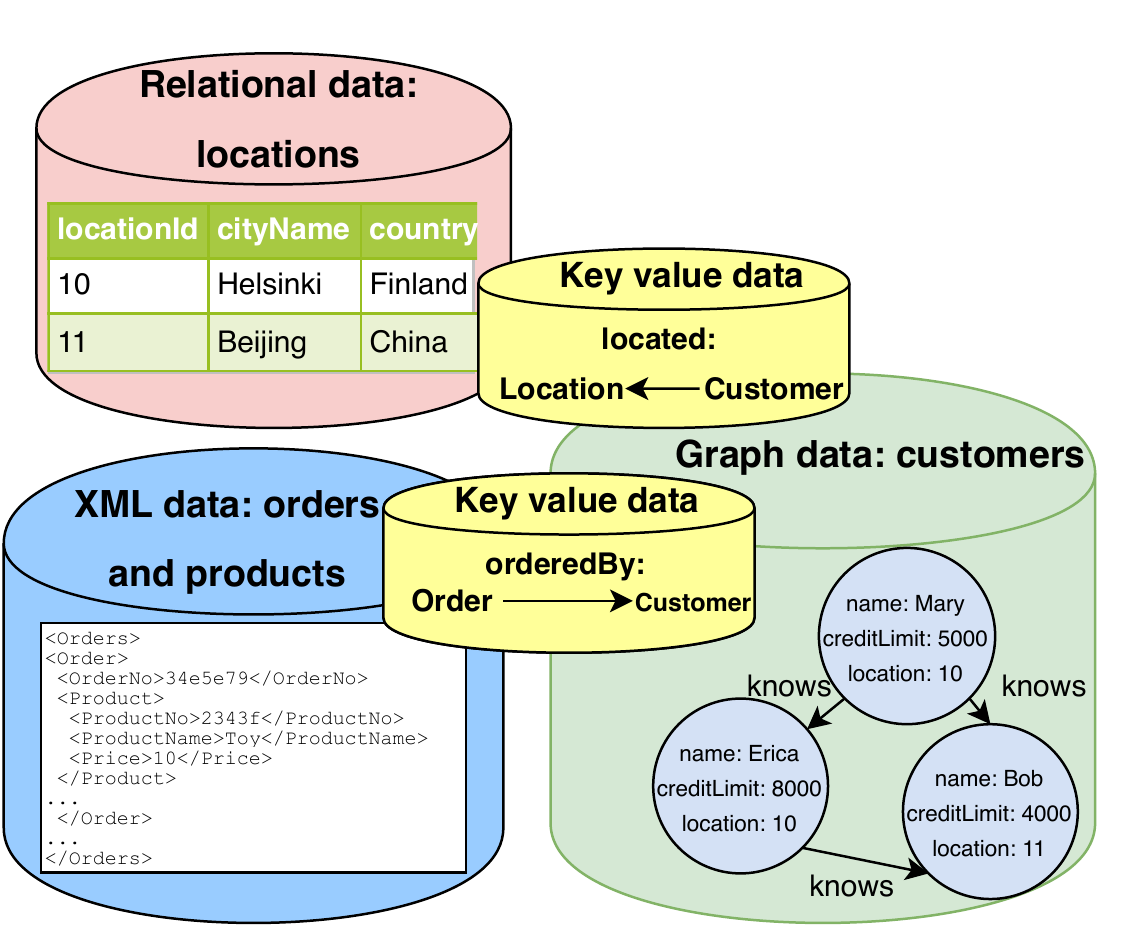}
    \caption{A multi-model data environment }
    \label{fig:dataSet}
\end{figure}

 Let us consider an example of a multi-model data environment. Figure \ref{fig:dataSet}  illustrates an application of E-commerce, which contains customers, a social network, and orders information with four distinct data models. The property graph data bear information about mutual relationships between the customers, i.e. who knows whom, and some customer properties such as name and credit limit. The geographic location of customers is stored in a  relational table. In XML documents, each order has an ID and a sequence of ordered products, each of which includes product number, name, and price. The fourth type of data, key/value pairs, contains the relations between different data sets. In a typical application like customer-360-view, users of databases demand to analyze the information from these four different data sources together to enable a holistic analysis of customer behaviors.
 
 
 Category theory was developed by mathematicians in the 1940s and has been successfully applied in many areas of science including computer science. Recent research initiatives have applied category theory for the database area. In particular, Spivak \cite{DBLP:journals/corr/abs-1009-1166, 10.5555/2628001} used a schema category and an instance functor to model relational databases. Liu et al. \cite{conf/vldb/LiuLGHPW18} promoted category theory to play the role of the new mathematical foundation to reason about declarative constructions and transformations between various data models.


While the previous works have envisioned the theoretical significance to model and manage data with category theory, this demonstration shows our initiative to showcase a proof-of-concept implementation of \texttt{Multi\-Category}, a system to support multi-model query processing based on category theory.  The core parts of the system have been coded with the functional programming language Haskell, which is widely recognized to have a strong connection to category theory. 
The data storing framework of \texttt{MultiCategory} is established on the concepts of schema and instance categories \cite{10.5555/2628001}, and the query processing structure is based on catamorphism and foldable data structures \cite{Grust1999Compr-709}. With these key properties, we can create a system that has a consistent integration with relational, hierarchical, and graph data models and we show how category theory can be used to achieve valuable perspectives for multi-model query representation and processing.

In brief, the demonstration of \texttt{MultiCategory} offers the following to the audience:
\begin{itemize}[noitemsep, topsep=0pt]
\item category theoretical and functional programming oriented methods of querying and accessing multi-model data with a unified schema;
\item a unified query language endowed with Haskell's lambda expressions allowing the users to submit one query to access different models of data  seamlessly;
\item the flexibility of output the same result with different models, which provides the users an opportunity to exploit the same data with different representations;
\item to better understand the theoretical structure behind the queries, this demo also provides an interactive visualizer to understand the schema and instance categories, as well as the query processing procedure.
\end{itemize}


In our demo, attendees are welcome to compose their queries that follow the syntax of our query language to search the multi-model datasets. The source code of this system is available in GitHub \cite{multicategoryDocs} and the demo video can be watched online on YouTube \cite{video}.

\section{Preliminaries}

In this section, we first review the mathematical definition of a category \cite{MacLane:205493}, followed by the descriptions of schema and instance categories which are influenced by \cite{DBLP:journals/corr/abs-1009-1166, 10.5555/2628001}.

\begin{definition}\label{def:category}
A category $\mathcal{C}$ consists of a collection of objects denoted by $Obj(\mathcal{C})$ and a collection of morphisms denoted by $Hom(\mathcal{C})$.  For each morphism $f \in Hom(\mathcal{C})$ there exists an object $A \in Obj(\mathcal{C})$ that is a domain of $f$ and an object $B \in Obj(\mathcal{C})$ that is a target of $f$. In this case we denote $f \colon A \to B$. We require that all the defined compositions of morphisms are included in $\mathcal{C}$: if $f\colon A \to B \in Hom(\mathcal{C})$ and $g \colon B \to C \in Hom(\mathcal{C})$, then $g \circ f \colon A \to C \in Hom(\mathcal{C})$. We assume that the composition operation is associative and that for every object $A \in Obj(\mathcal{C})$ there exists an identity morphism $\text{id}_{A} \colon A \to A$ so that $f \circ \text{id}_{A} = f$ and $\text{id}_{A} \circ f = f$ whenever the composition is defined.
\end{definition}

Informally speaking, we can understand a category as a graph endowed with the composition rule.


 \begin{figure}
    \centering
    \includegraphics[scale=0.4]{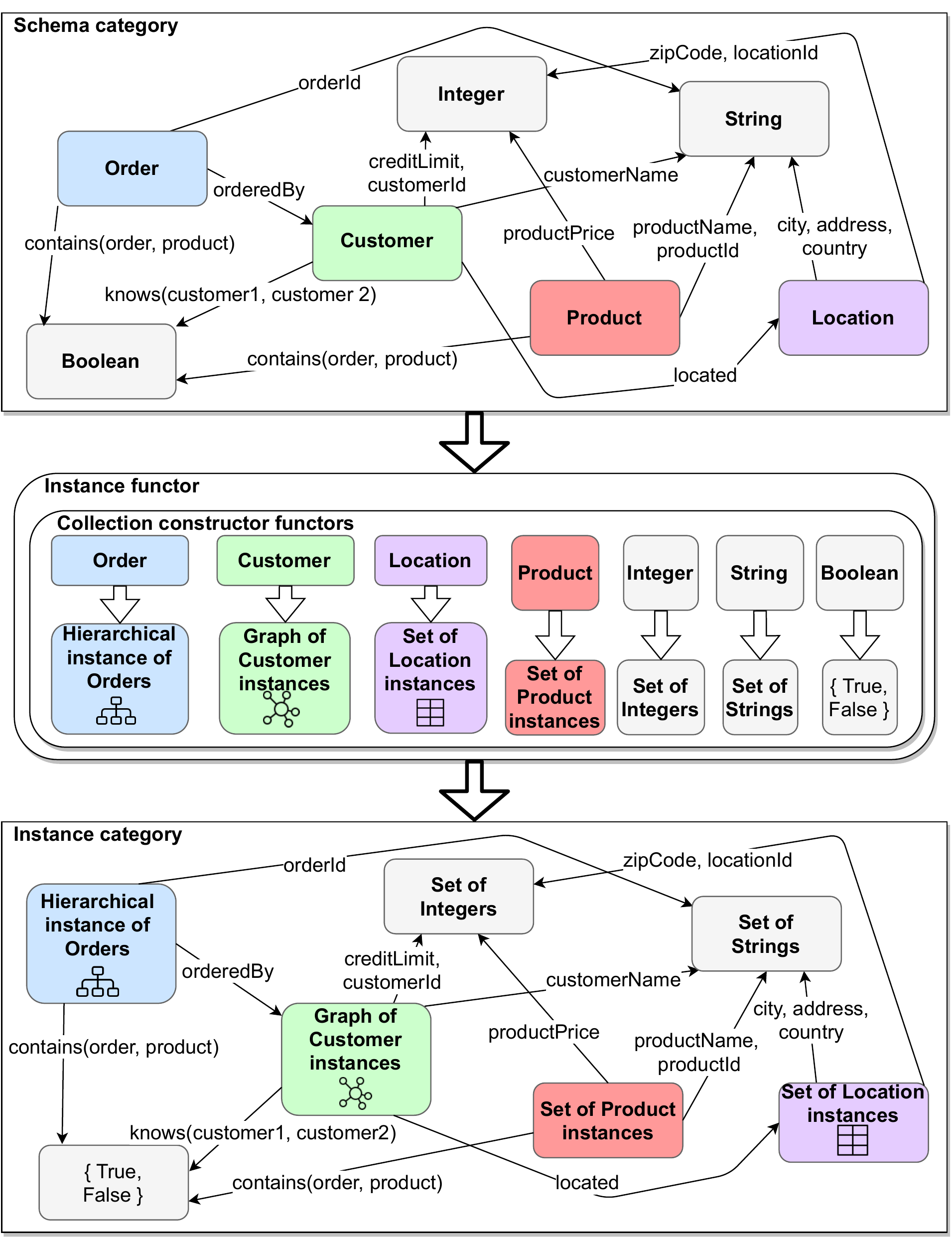}
    \caption{An example of a category theoretical construction}
    \label{fig:schema_category_example}
\end{figure}

Figure \ref{fig:schema_category_example} constructs a unified schema category, which represents the schema information of a multi-model data environment in Figure \ref{fig:dataSet}.  Conceptually, an object in a schema category includes two kinds of data types: (1) the first collection of data types consists of a string, integer, rational, boolean, etc., called predefined data types; and (2) the second collection of data types includes entities, such as customers and products. Morphisms are defined to be the typed functions between the data types, such as a customer is located in a certain location and an order is ordered by a customer. Furthermore, it is important to note that a schema category presents a single unified view for different models of data. Based on this view, we develop a unified query mechanism to process different models of data seamlessly.


An instance category models how the concrete data instances are stored. Each object of the schema category is mapped to the corresponding typed Haskell data structure in the instance category (see Figure \ref{fig:schema_category_example}). Each morphism in the schema category is mapped to a concrete Haskell function in the instance category. The mapping between these categories is called an \textit{instance functor} which is defined on objects by collection constructor functors \cite{Grust1999Compr-709}.  As shown in our demo, queries are formulated based on the schema category and the answers are retrieved from the instance category based on the instance functor and the collection constructor functors.


In schema and instance categories, we can follow any path to form a well-defined function between the start node and the end node of the path thanks to the composition rule in Definition \ref{def:category}. For example, there is a morphism (edge) in the instance category that gives us that the customer $A$ makes the order $B$ and another morphism that gives us that the order $B$ includes the product $C$. Based on the composition rule of these morphisms there is a well-defined morphism that gives us that the customer $A$ buys the product $C$. This compositionality property is important to guarantee the correctness of programs to traverse through multiple data models.

\section{System overview}
In this section, we provide an overview of \texttt{MultiCategory}'s architecture, query language, and query processing mechanism. For more details about technical solutions, a tutorial, and an installation guide you can find from \texttt{MultiCategory}'s documentation and in Github \cite{multicategoryDocs}.

\begin{figure}   
    \centering
    \includegraphics[scale=0.56]{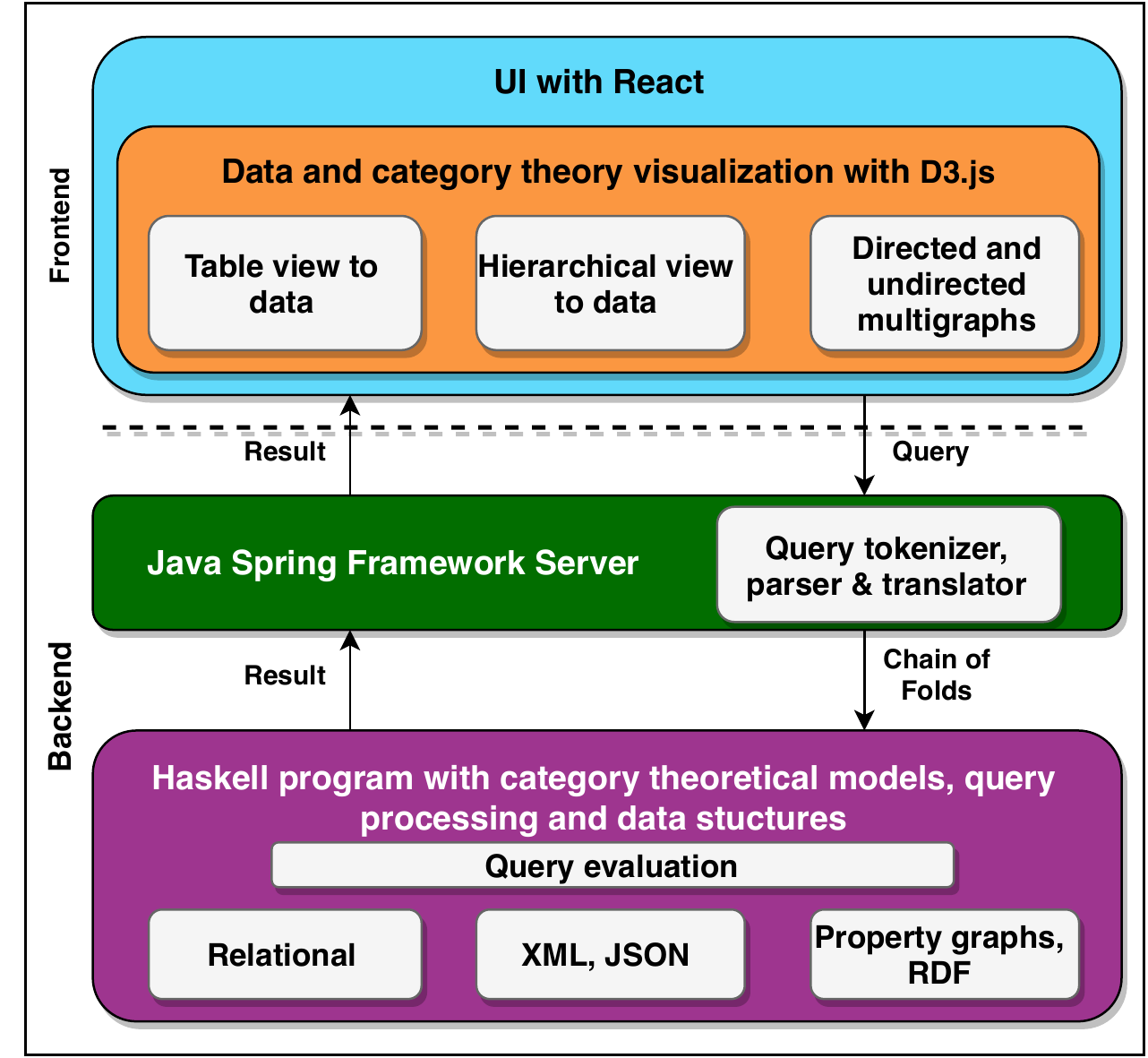} 
    \caption{The architecture of \texttt{MultiCategory}} 
    \label{fig:architecture}
\end{figure}

Figure \ref{fig:architecture} depicts the architecture of \texttt{MultiCategory} that consists of the frontend and the backend. In particular, the frontend creates a web interface and data visualizations for relational data, hierarchical documents, and graphs. The backend is responsible for query processing and the implementation of category theoretical constructions. 

\subsection{Multi-model query language}

We have developed a  multi-model query language that encapsulates Haskell functions and expressions. For example, the following query finds \textit{those customers whose credit limit is greater than a threshold, say 3000}.

\begin{example}\label{ex:simpleExample1}
\begin{fslisting} 
 QUERY (\x -> if creditLimit x > 3000 then cons x else nil)
FROM customers
TO graph/xml/relational
\end{fslisting}
\end{example}

\begin{figure*}[ht]
   \begin{minipage}{0.33\textwidth}
    \centering
    \includegraphics[width=\linewidth, height = 3cm]{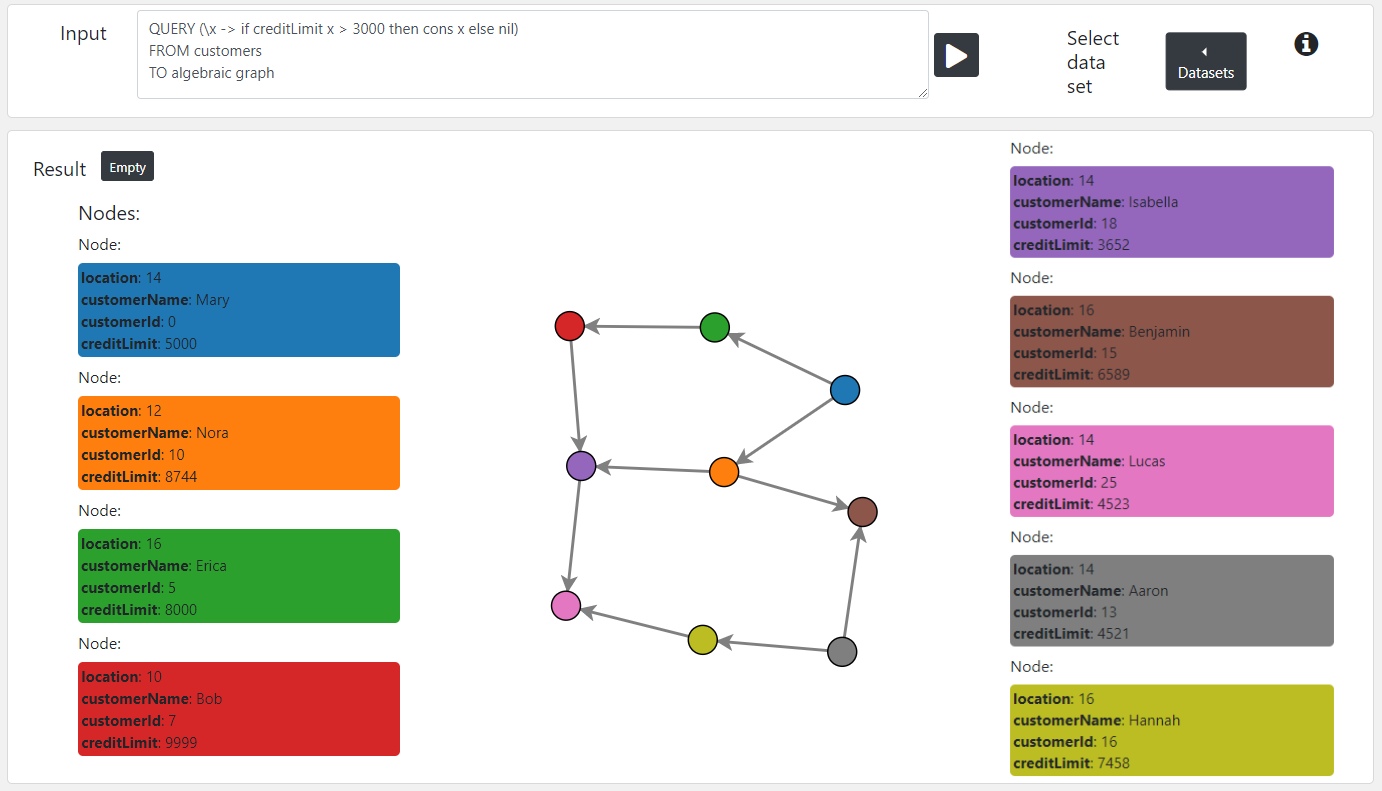}
    \subcaption{Graph}
   \end{minipage}\hfill
   \begin{minipage}{0.33\textwidth}
     \centering
    \includegraphics[width=\linewidth, height = 3cm]{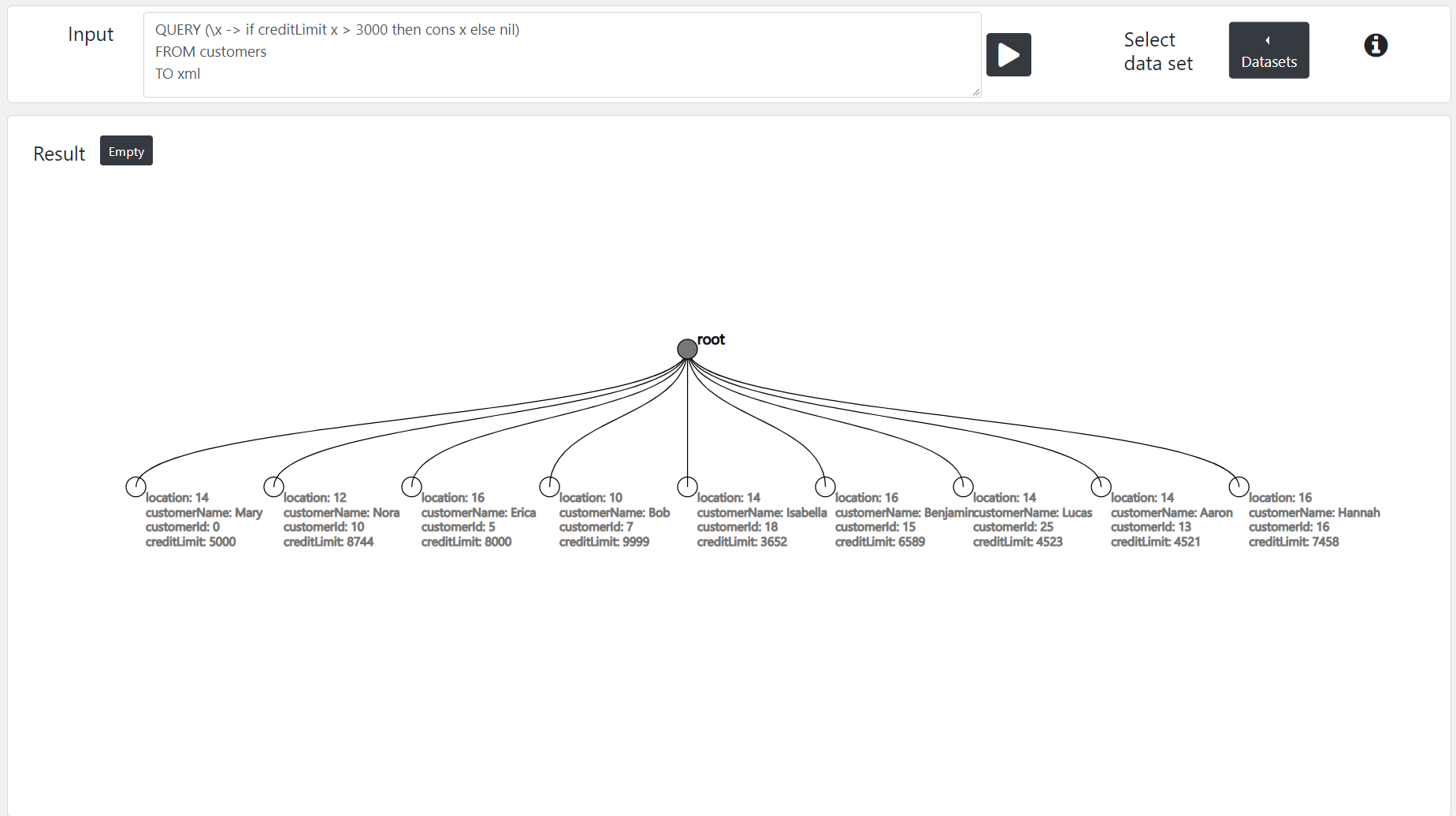}
    \subcaption{Tree}
   \end{minipage}\hfill
   \begin{minipage}{0.33\textwidth}
     \centering
    \includegraphics[width=\linewidth, height = 3cm]{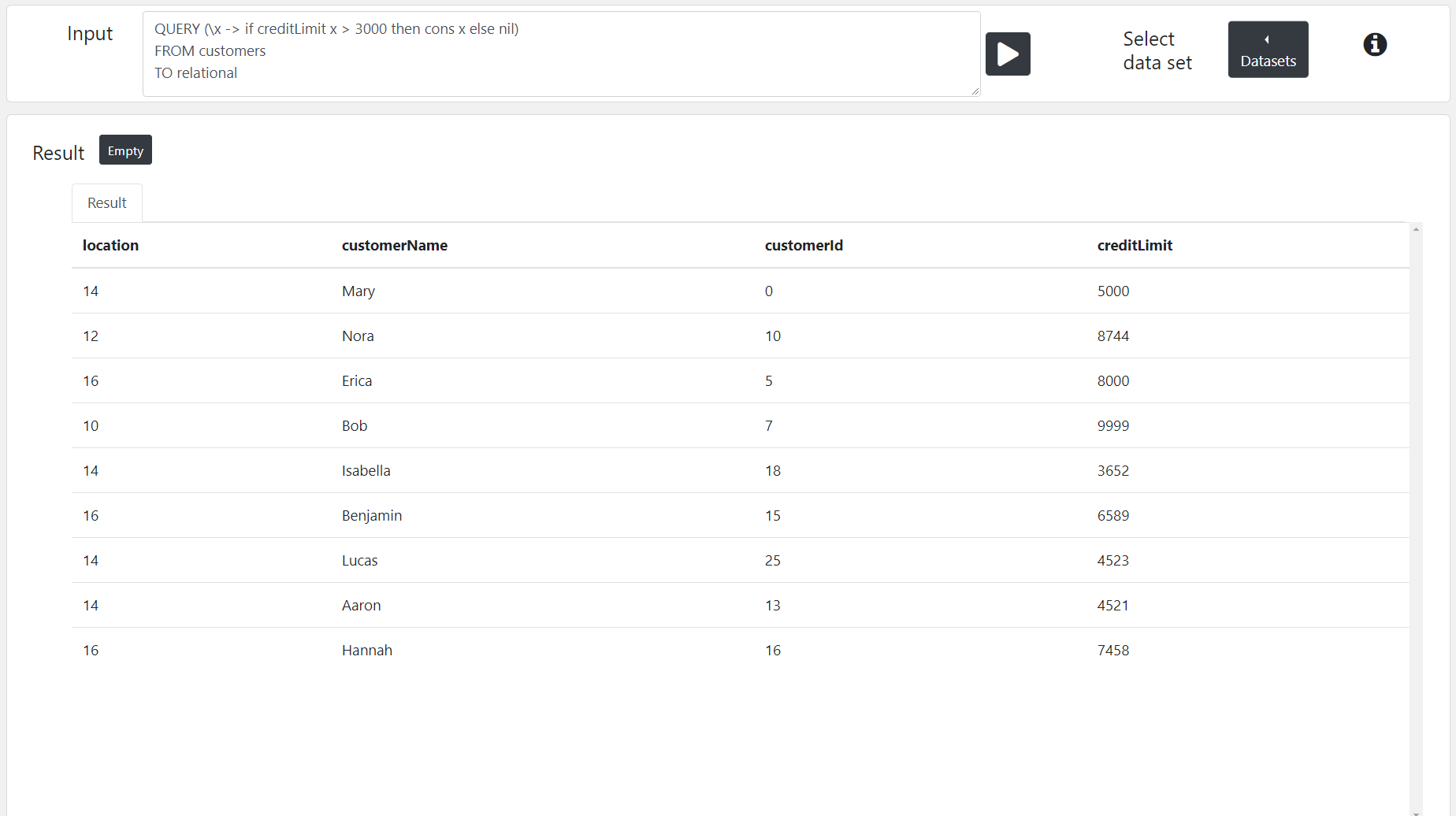}
    \subcaption{Relation}
   \end{minipage}
   \caption{Three different representations of the same query result of Example \ref{ex:simpleExample1}}
   \label{fig:simpleExampleResult}
\end{figure*}

Every query block starts with \textbf{QUERY} keyword, which is followed by a function defined with Haskell's lambda notation. In Example \ref{ex:simpleExample1}, \texttt{x} is a variable that represents a customer in the graph, and the query returns the graph of the customers with credit limit $> 3000$.  The \textbf{FROM} keyword specifies the source collection of the data.  The \textbf{TO} keyword configures the data model of the result. Note that in the above example the returned model can be either \textit{graph}, \textit{relational} or \textit{XML}. Figure \ref{fig:simpleExampleResult} demonstrates the three different representation models for the answers of the query in Example \ref{ex:simpleExample1}.

\begin{example}\label{ex:longExample1}
\begin{fslisting}
LET t BE
QUERY (\x xs -> if elem "Book" (map productName (orderProducts x)) then cons x xs else xs)
FROM orders TO relational IN
QUERY (\x -> if any (\y -> orderedBy y customers == x) t then cons (customerName x, countryName(located x locations)) else nil)
FROM customers TO algebraic graph/relational/xml
 \end{fslisting}
 \end{example}

\begin{figure}
    \centering
    \includegraphics[scale=0.30]{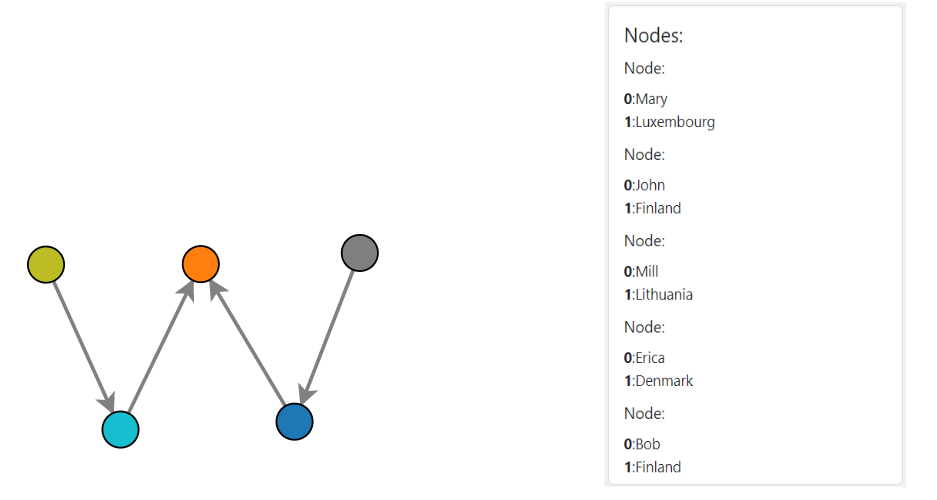}
    \includegraphics[height = 2.2cm, width=\linewidth]{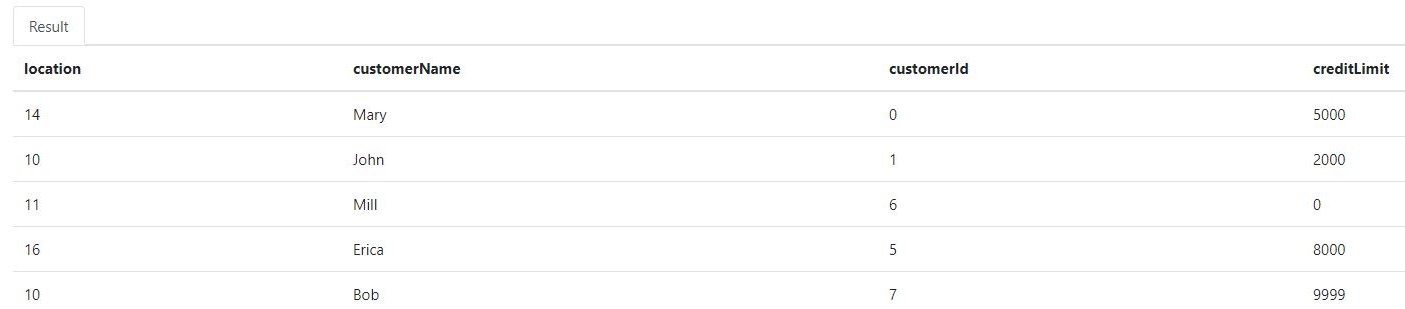}
    \caption{The result in graph and relational models for the query in Example \ref{ex:longExample1}}
    \label{fig:example3}
\end{figure}

The query in Example \ref{ex:longExample1} returns a graph that contains names and locations of the customers who ordered a book, which involves relation, XML, and key/value data types. Figure \ref{fig:example3} shows the result of this query. There are two QUERY clauses in this query that correspond to the two fold functions. The clauses are connected with \textbf{LET BE IN} structure which works in the same way as the corresponding mechanism in Haskell. The \textbf{LET} keyword introduces a variable (i.e. $t$) that connects fold functions together. In particular,  the first \textbf{QUERY} clause finds any order which contains a book. The second \textbf{QUERY} clause finds customers who made such orders. The results contain the customer's name and location information (i.e. countryName).

\subsection{Query processing mechanism}

\begin{figure}
    \centering
    \includegraphics[scale = 0.5]{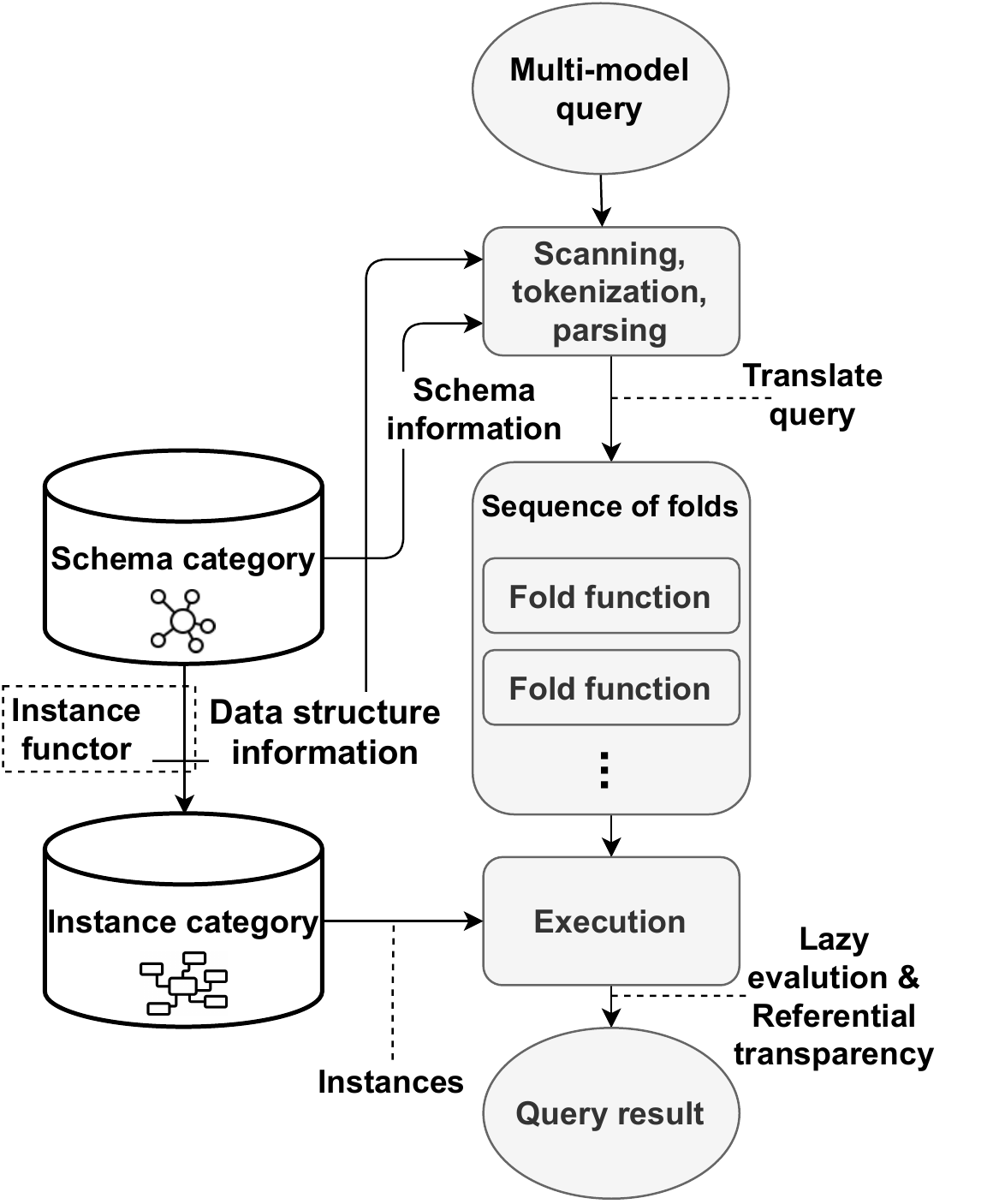}
    \caption{The workflow diagram of query processing}
    \label{fig:queryExecution}
\end{figure}

Figure \ref{fig:queryExecution} depicts the main workflow of query processing in \texttt{Multi\-Category}. When a user inputs a query to the system, it is parsed into a sequence of \textit{fold-functions} with respect to the schema information from the schema category. The sequence of folds is sent to the Haskell program running in the backend. The Haskell program accesses the instance category and executes the sequence of fold functions that is just pure Haskell code. Note that we do not require all data structures to be instances of Haskell's foldable type class as we can use generalizations of folds to query more complex algebraic data types. For example, we use the function \texttt{foldg} to query the algebraic graphs of the package \texttt{Algebra.\-Graph} \cite{10.1145/3122955.3122956}. Finally, the result is returned to the frontend, where it is visualized depending on the model the user defined in the query. 



\section{Demonstration scenarios}


\subsection{Using schema and instance categories}

We first introduce our system by inviting attendees to view the schema and instance categories of varied data sets through the graphical interface. This demo will use six different synthetic and real datasets. Each data set includes different models, such as relational, XML, JSON, RDF, and property graph data. Attendees can select a data set, view the related schema and instance categories, and examine the nodes and edges of any graph to find more information. For example, one may find that \textit{customer} data type has an attribute \textit{customerName} which is considered as a function from \texttt{Customer} to \texttt{String}. 

\subsection{Querying multi-model data}
In \texttt{MultiCategory} we have created a collection of example queries that can query different data models together. For example, an E-commerce data set includes all the seven possible combinations of multi-model queries combining relational, XML, and graph models. This demo allows attendees to formulate their queries with guidance and observe results in different output models. Currently, the system does not support large-scale data processing due to implementational reasons but generally category theory-based frameworks scale well.

\subsection{Visualizing multi-model queries}

For obtaining a better understanding of the theoretical structure behind the queries, \texttt{MultiCategory} provides an automatic visualizer for the fold function-based queries. This feature visualizes queries as graphs with respect to the instance category. In particular, after the execution of a query, the query is visualized as a graph that exhibits a composition of Haskell fold functions. Every fold function has at least one lambda expression which the user can click to see the detailed information in the graphical interface. 

\section{Comparison to existing systems}

The existing multi-model databases, for example, ArangoDB and OrientDB are implemented based on a single dominant model so that they cannot be called ``native'' concerning all models they are supporting.  
In contrast, we developed \texttt{MultiCategory} in such a way that each data model is equal and each instance is stored in its native data structure. No specific transformation operations are required when the data is uploaded. A unified view is generated to accommodate different models of data together. Besides, there are very few systems that would support as many models (relational, XML, JSON, RDF, and property graphs) as \texttt{MultiCategory} supports. 

\section{Conclusion and future work}

\texttt{MultiCategory} is a tangible system that is applying category theory to model and query multi-model data. We implement a  query language with a  functional programming language.  We visualize the category theoretical constructions, i.e. schema and instance categories, and query processing to show connections between theory and applications.  Note that \texttt{MultiCategory} is not yet a full-fledged database system and due to its implementation it is also not a big data system. 

The schema category and instance category are fixed and predefined manually. In the future, we consider automatically generating this unified category view based on input data sets. We have been developing the theoretical framework so that it can define multi-model joins and data, schema, and query transformations. 


\bibliographystyle{ACM-Reference-Format}
\bibliography{main}


\begin{thebibliography}{9}


\ifx \showCODEN    \undefined \def \showCODEN     #1{\unskip}     \fi
\ifx \showDOI      \undefined \def \showDOI       #1{#1}\fi
\ifx \showISBNx    \undefined \def \showISBNx     #1{\unskip}     \fi
\ifx \showISBNxiii \undefined \def \showISBNxiii  #1{\unskip}     \fi
\ifx \showISSN     \undefined \def \showISSN      #1{\unskip}     \fi
\ifx \showLCCN     \undefined \def \showLCCN      #1{\unskip}     \fi
\ifx \shownote     \undefined \def \shownote      #1{#1}          \fi
\ifx \showarticletitle \undefined \def \showarticletitle #1{#1}   \fi
\ifx \showURL      \undefined \def \showURL       {\relax}        \fi
\providecommand\bibfield[2]{#2}
\providecommand\bibinfo[2]{#2}
\providecommand\natexlab[1]{#1}
\providecommand\showeprint[2][]{arXiv:#2}

\bibitem[\protect\citeauthoryear{Grust}{Grust}{1999}]%
        {Grust1999Compr-709}
\bibfield{author}{\bibinfo{person}{Torsten Grust}.}
  \bibinfo{year}{1999}\natexlab{}.
\newblock \emph{\bibinfo{title}{Comprehending Queries}}.
\newblock \bibinfo{thesistype}{Ph.D. Dissertation}.
  \bibinfo{school}{Universitaet Konstanz}, \bibinfo{address}{Konstanz}.
\newblock


\bibitem[\protect\citeauthoryear{Liu, Lu, Gawlick, Helskyaho, Pogossiants, and
  Wu}{Liu et~al\mbox{.}}{2018}]%
        {conf/vldb/LiuLGHPW18}
\bibfield{author}{\bibinfo{person}{Zhen~Hua Liu}, \bibinfo{person}{Jiaheng Lu},
  \bibinfo{person}{Dieter Gawlick}, \bibinfo{person}{Heli Helskyaho},
  \bibinfo{person}{Gregory Pogossiants}, {and} \bibinfo{person}{Zhe Wu}.}
  \bibinfo{year}{2018}\natexlab{}.
\newblock \showarticletitle{Multi-model Database Management Systems - {A} Look
  Forward}. In \bibinfo{booktitle}{\emph{Polystores {VLDB} 2018 Workshops}}.
  \bibinfo{pages}{16--29}.
\newblock


\bibitem[\protect\citeauthoryear{Lu and Holubov{\'{a}}}{Lu and
  Holubov{\'{a}}}{2019}]%
        {journals/csur/LuH19}
\bibfield{author}{\bibinfo{person}{Jiaheng Lu} {and} \bibinfo{person}{Irena
  Holubov{\'{a}}}.} \bibinfo{year}{2019}\natexlab{}.
\newblock \showarticletitle{Multi-model Databases: {A} New Journey to Handle
  the Variety of Data}.
\newblock \bibinfo{journal}{\emph{{ACM} Comput. Surv.}} \bibinfo{volume}{52},
  \bibinfo{number}{3} (\bibinfo{year}{2019}), \bibinfo{pages}{55:1--55:38}.
\newblock


\bibitem[\protect\citeauthoryear{MacLane}{MacLane}{1971}]%
        {MacLane:205493}
\bibfield{author}{\bibinfo{person}{Saunders MacLane}.}
  \bibinfo{year}{1971}\natexlab{}.
\newblock \bibinfo{booktitle}{\emph{{Categories for the working
  mathematician}}}.
\newblock \bibinfo{publisher}{Springer}, \bibinfo{address}{New York, NY}.
\newblock
\urldef\tempurl%
\url{https://doi.org/10.1007/978-1-4612-9839-7}
\showDOI{\tempurl}


\bibitem[\protect\citeauthoryear{Mokhov}{Mokhov}{2017}]%
        {10.1145/3122955.3122956}
\bibfield{author}{\bibinfo{person}{Andrey Mokhov}.}
  \bibinfo{year}{2017}\natexlab{}.
\newblock \showarticletitle{Algebraic Graphs with Class (Functional Pearl)}. In
  \bibinfo{booktitle}{\emph{Proceedings of the 10th ACM SIGPLAN International
  Symposium on Haskell}} (Oxford, UK) \emph{(\bibinfo{series}{Haskell 2017})}.
  \bibinfo{address}{New York, NY, USA}, \bibinfo{pages}{2–13}.
\newblock
\showISBNx{9781450351829}
\urldef\tempurl%
\url{https://doi.org/10.1145/3122955.3122956}
\showDOI{\tempurl}


\bibitem[\protect\citeauthoryear{Spivak}{Spivak}{2014}]%
        {10.5555/2628001}
\bibfield{author}{\bibinfo{person}{David Spivak}.}
  \bibinfo{year}{2014}\natexlab{}.
\newblock \showarticletitle{Category Theory for the Sciences}.
\newblock  (\bibinfo{year}{2014}).
\newblock
\showISBNx{0262028131}


\bibitem[\protect\citeauthoryear{Spivak}{Spivak}{2010}]%
        {DBLP:journals/corr/abs-1009-1166}
\bibfield{author}{\bibinfo{person}{David~I. Spivak}.}
  \bibinfo{year}{2010}\natexlab{}.
\newblock \showarticletitle{Functorial Data Migration}.
\newblock \bibinfo{journal}{\emph{CoRR}}  \bibinfo{volume}{abs/1009.1166}
  (\bibinfo{year}{2010}).
\newblock
\showeprint[arxiv]{1009.1166}
\urldef\tempurl%
\url{http://arxiv.org/abs/1009.1166}
\showURL{%
\tempurl}


\bibitem[\protect\citeauthoryear{Uotila}{Uotila}{2021}]%
        {multicategoryDocs}
\bibfield{author}{\bibinfo{person}{Valter Uotila}.}
  \bibinfo{year}{2021}\natexlab{}.
\newblock \bibinfo{title}{MultiCategory Documentation and System Codes}.
\newblock \bibinfo{howpublished}{\url{https://multicategory.github.io/},
  \url{https://git.io/JvPqM}}.
\newblock
\newblock
\shownote{Accessed Jul. 19, 2021.}


\bibitem[\protect\citeauthoryear{Uotila and Lu}{Uotila and Lu}{2021}]%
        {video}
\bibfield{author}{\bibinfo{person}{Valter Uotila} {and}
  \bibinfo{person}{Jiaheng Lu}.} \bibinfo{year}{2021}\natexlab{}.
\newblock \bibinfo{title}{{MultiCategory} demo video}.
\newblock \bibinfo{howpublished}{\url{https://youtu.be/uceIi91AGsg}}.
\newblock
\newblock
\shownote{Accessed Jul. 19, 2021.}


\end{thebibliography}

\end{document}